\documentclass[12pt]{article}
\usepackage{amsfonts}
\usepackage{amsmath}
\usepackage{amssymb}
\usepackage{graphicx}
\usepackage{float}
\usepackage{color}
\usepackage{epigraph}
\usepackage{setspace} 
\usepackage{physics}
\usepackage{breqn}
\textwidth=6.5in \hoffset=-.75in \voffset=-1in \numberwithin{equation}{section}
\numberwithin{figure}{section}

\newcommand {\nn}{\nonumber}
\newcommand {\be}{\begin{equation}}
\newcommand {\ee}{\end{equation}}

\begin{document}

\begin{titlepage}
\vspace{1cm}
\begin{center}
{\Large \bf {On the relationship between charge and mass in General Relativity}}\\
\end{center}
\vspace{2cm}
\begin{center}
{Vineet Kumar \footnote{E-Mail: v.kumar@usask.ca}}
\vspace{1cm}
\vspace{2cm}
\end{center}

\begin{abstract}
This paper demonstrates a relationship between mass and charge through explicit construction of exact Einstein-Maxwell spacetimes by embedding the Schwarzschild and Kerr instantons in 5 dimensions. It is shown further how by varying only the mass of an object, its electric field or apparent ``charge'' changes, while making it possible for two different mass values to yield the same electric field strength.
\end{abstract}
\end{titlepage}\onecolumn 
\bigskip 
\epigraph{\singlespace{Study hard what interests you the most in the most undisciplined, irreverent and original manner possible.}}%
{\textsc{Richard Feynman}}

We construct a new exact solution to Einstein-Maxwell theory by embedding the Kerr instanton \cite{Gib77} in 5-dimensional Einstein-Maxwell theory and explicitly show that the same parameter (``mass''\footnote{The quotation marks are not preposterous; Rainich-Misner-Wheeler geometrodynamics even offers ``\textit{mass without mass}''.}) that determines the gravitational field strength also determines the electric field strength, thereby questioning the notion of electric ``charge'' as a property of matter independent of mass. Furthermore, we also show for any given electric field configuration, there are two possible mass values that yield it.

The idea of electric charge as an illusion goes back to at least as far as John Wheeler who referred to the chargelike property of wormholes as ``charge without charge''. In their classic treatise on ``{\it geometrodynamics}'', John Wheeler and Charles Misner \cite{Mis57} do a thorough exploration of Rainich's work \cite{Rai25} on the geometrisation of Einstein and Maxwell equations and illustrate how electric field lines may emerge out of ``holes'' or handles in multiply connected space but appear to a macroscopic observer as emerging from a point source in simply connected space and thus making him erroneously apply Gauss's law on a perceived ``point charge''.

\section{Charge from Mass}
To illustrate the idea, we begin with an exact solution of 5-dimensional Einstein-Maxwell equations. The trace-free\footnote{Because $T_{\mu\nu}$ is trace-free, the Ricci scalar vanishes.} form of source-free Einstein-Maxwell equations in $D$ dimensions in geometrised units are given by:
\begin{eqnarray}
R_{\mu\nu}&=&F_{\mu}^{\lambda}F_{\nu\lambda}-\frac{1}{4+2(D-4)}g_{\mu\nu}F^2\label{EinMax},\\
F^{\mu\nu}_{;\mu}&=&0\label{Max}
\end{eqnarray}
Our solution is constructed by the embedding of a 4-dimensional gravitational instanton.\cite{GK2017_1}\cite{GK2017_2} Recall that the Kerr blackhole has only mass and angular momentum terms and no charge term. We get the Kerr instanton from Kerr blackhole metric through appropriate Wick rotations. 

The metric for the Kerr instanton \cite{Gib77} is given by the line element \cite{Gro82}:
\begin{dmath}
ds_{KerrInst}^2=(r^2-a^2 \cos^2 \theta)\left(\frac{d r^2}{\Delta (r)}
+d \theta^2\right)+\frac{1}{r^2-a^2 \cos^2 \theta}(\Delta (r) (d \tau + a \sin ^2 \theta \,d\phi)^2 +\sin^2 \theta ((r^2-a^2)d\phi -a \,{d}\tau)^2)
\end{dmath}
where $\Delta (r)=r^2-r_s r-a^2$,\,\, $a=\dfrac{\cal J}{Mc}$ and $r_s=\dfrac{2GM}{c^2}$.

For $a=0$, i.e., ${\cal J}=0$, the Kerr instanton becomes the Schwarzschild instanton \cite{Gro82}. The coordinates $r,\,\theta,\,\phi$ are the spherical coordinates and the coordinate $\tau$ is obtained from the Kerr spacetime by a Wick rotation of the timelike coordinate via $t\rightarrow i \tau$. The new spacelike coordinate $\tau$ must be periodic \cite{Gib77}.

Then one proceeds with the following embedding ansatz for five dimensions:\footnote{I have included $c^2$ in the metric to aid the analysis ahead although it is not necessary.}
\be
ds^2=-c^2\frac{dt^2}{H^2(r)}+H(r)\,ds_{KerrInst}^2
\ee

The only non-zero component of the gauge field is given by:\footnote{At this point, it should be noted that the Kaluza ``cylinder condition'' is tacitly assumed here, i.e., that the metric and the gauge function are independent of the coordinate $\tau$.}
\be
A_t=\sqrt{\frac{3}{2}}\frac{1}{H(r)}
\ee
Einstein and Maxwell equations, in order to be satisfied, demand that the metric function $H(r)$ satisfies the following ODE:
\be
(-c^2r^2+2GMr+a^2c^2)\frac{d^2 H(r)}{d r^2}+(2GM-2c^2 r)\frac{d H(r)}{d r}=0
\label{kerr5dode}
\ee

\begin{figure}[H]
	\centering
	\includegraphics[width=0.45\textwidth]{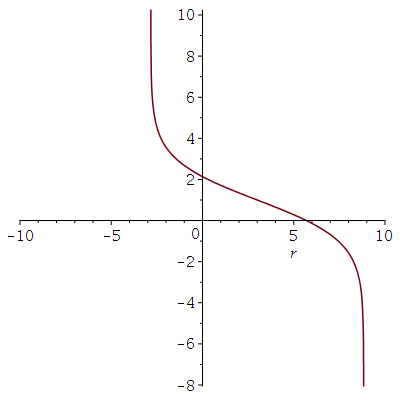}
	\includegraphics[width=0.45\textwidth]{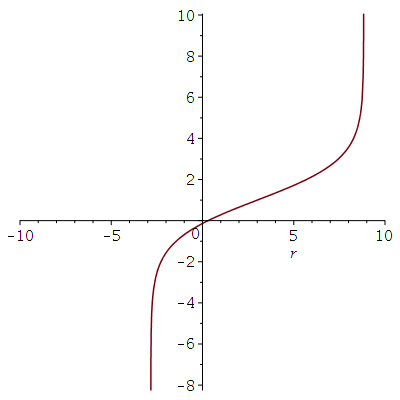}
	\caption{Plots of $H_{+}(r)$ with  $c=1,\,c_1=1,\,\,G=1,\,M=3,\,a=5,\,C_2=2$ (left) and $C_2=-2$ (right)}
	\label{arctanhHr}
\end{figure}

A nice, healthy solution to the above equation is:
\be
H_{+}(r)=c_1+C_2 \tanh^{-1} \left( \frac{GM-c^2 r}{\sqrt{a^2 c^4+G^2 M^2}}\right)
\ee

where $c_1$ and $C_2$ are constants. Figure \ref{arctanhHr} shows a plot of the function $H(r)$.

Looks like the solution $H_{+}(r)$ suffers from a pathology as it is not defined for $ r>\frac{{GM}-{\sqrt{a^2 c^4+G^2 M^2}}}{c^2}$.

\begin{figure}[H]
	\centering
	\includegraphics[width=0.4\textwidth]{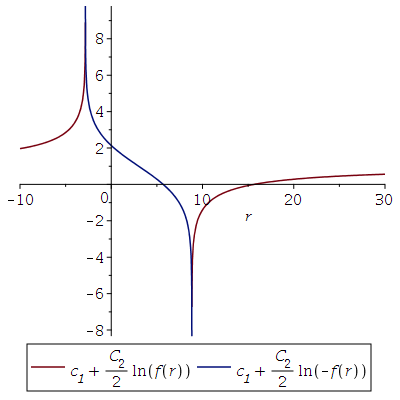}
	\includegraphics[width=0.4\textwidth]{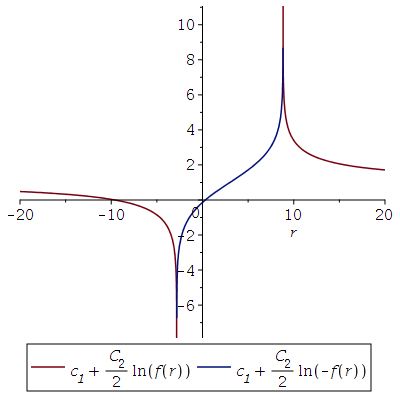}
	\caption{Plots of $H_{ext}(r)$ and $H_{int}(r)$ with  $c=1,\,c_1=1,\,\,G=1,\,M=3,\,a=5,\,C_2=2$ (left) and $C_2=-2$ (right)}
	\label{Hlnplot}
\end{figure}

Evidently, the pathology is not easily cured by a sign change but we should not give up and instead convert $\tanh^{-1}$ to its logarithmic form using:
\be
\tanh^{-1}(x)=\frac{1}{2}\ln\frac{1+x}{1-x}
\ee
Therefore,
\begin{align}
H_{ext}(r)
&=c_1+\frac{C_2}{2} \ln\left(\dfrac{1+\dfrac{GM-c^2 r}{\sqrt{a^2 c^4+G^2 M^2}}}{1-\dfrac{GM-c^2 r}{\sqrt{a^2 c^4+G^2 M^2}}}\right)\\
&=c_1+\frac{C_2}{2} \ln\left(\dfrac{{\sqrt{a^2 c^4+G^2 M^2}}+{GM-c^2 r}}{{\sqrt{a^2 c^4+G^2 M^2}}-{GM}+{c^2 r}}\right)\\
&=c_1+\frac{C_2}{2} \ln(f(r))
\end{align}
is a solution to Eq.~(\ref{kerr5dode}), with $f(r)$ representing the argument of the logarithmic function. Clearly, $H_{ext}(r)$ is not valid for $f(r)<0$. Thus, we might as well write another solution $H_{int}(r)=c_1+\frac{C_2}{2} \ln(-f(r))$. Figure \ref{Hlnplot} shows plots of the two functions for some arbitrary values of the constants.
Ultimately, the most wholesome solution, that comprises of both $H_{ext}(r)$ and $H_{int}(r)$, can be written as:\footnote{If you find the absolute value distasteful but are fond of even integers, then writing $H(r)=c_1+\frac{1}{4}\ln\left(\left({\dfrac{{\sqrt{a^2 c^4+G^2 M^2}}+{GM-c^2 r}}{{\sqrt{a^2 c^4+G^2 M^2}}-{GM}+{c^2 r}}}\right)^{c_2}\right)$ and demanding that $c_2$ be an even integer, should assuage your immediate concern but it may raise a new question\textemdash ``why integers?!''}
\begin{align}
H(r)&=c_1+\dfrac{c_2}{4} \ln(|f(r)|)\nn\\
&=c_1+\dfrac{c_2}{4}\ln\left(\lvert{\dfrac{{\sqrt{a^2 c^4+G^2 M^2}}+{GM-c^2 r}}{{\sqrt{a^2 c^4+G^2 M^2}}-{GM}+{c^2 r}}}\rvert\right)
\end{align}
where $c_2=2\,C_2$.
\section{Event horizons}
The event horizons are located where $g_{tt}$ vanishes:
\be
g_{tt}=-\frac{c^2}{\left(c_1+\dfrac{c_2}{4}\ln\left(\abs*{\dfrac{{\sqrt{a^2 c^4+G^2 M^2}}+{GM-c^2 r}}{{\sqrt{a^2 c^4+G^2 M^2}}-{GM}+{c^2 r}}}\right)\right)^2}=0
\ee
This implies that the logarithm function needs to approach infinity. The only possible location for $g_{tt}$ to vanish for $r>0$ then is:
\be
\Longrightarrow r_{eh}=\frac{\sqrt{a^2 c^4+G^2 M^2}+GM}{c^2}
\ee
which, of course, implies that in the limit $a\rightarrow0$, $r_{eh}$ becomes the Schwarzschild radius $\frac{2GM}{c^2}$.

\section{Weak field limit: Linearised gravity}
To find the strength of the gravitational field at infinity, we expand $g_{tt}$ as series and take the linear term in $\frac{1}{r}$:
\be
g_{tt}=-\frac{c^2}{c_1^2}-c_2 \frac{\sqrt{a^2 c^4+G^2 M^2}}{c_1^3 r}+O\left(\frac{1}{r^2}\right)
\label{g_tt}
\ee
In the weak field limit for a spherically symmetric, non-rotating, uncharged, gravitating body, we expect:
\be
g_{tt}=-c^2+\frac{2GM}{r}
\ee
Clearly, $c_1=\pm 1$. And $c_2$ can be fixed\footnote{The fixing of $c_2$ may not be as simple as that of $c_1$ and here, I have opted for the method motivated by a desire to model Newtonian gravity in the weak field limit.} by noting that for $a=0$ (i.e., the embedded Schwarzschild instanton where there is no coupling between $\tau$ and $r$ coordinates), we should have $c_2=\mp 2$ in order to model the attractive gravitational field of a non-rotating, spherically symmetric body.

Thus, the metric function $H(r)$ is given by:\footnote{Now that $c_2=\mp 2$ is an even integer, the universe does not have to compute an unnecessary {\it if-else} conditional arising from the absolute value.}
\be
H(r)=c_1+\dfrac{1}{4}\ln\left(\left({\dfrac{{\sqrt{a^2 c^4+G^2 M^2}}+{GM-c^2 r}}{{\sqrt{a^2 c^4+G^2 M^2}}-{GM}+{c^2 r}}}\right)^{c_2}\right)
\ee

where $c_1=\pm 1$ and $c_2=\mp 2$.

\section{Electric field}
Recall that the $r$-component (the only component, in this case) of the electric field is given by:
\begin{align}
E_r(r)=F_{tr}&=-\sqrt{\frac{3}{2}}\frac{1}{H^2(r)}\frac{d  H(r)}{dr}\\
&=-\frac{c_2}{2}\sqrt{\frac{3}{2}}\frac{1}{H^2(r)}\frac{\sqrt{a^2c^4+G^2M^2}}{(c^2r^2-a^2c^2-2GMr)}
\label{elecgen}
\end{align}

which can be expanded as series at infinity as:
\begin{align}
E_r(r)&=-\frac{\sqrt{3}}{2\sqrt{2}}\frac{c_2}{c^2 c_1^2}\frac{\sqrt{a^2c^4+G^2 M^2}}{r^2}\\
&=-c_2\frac{\sqrt{3}}{2\sqrt{2}}\frac{1}{c^2 }\frac{\sqrt{a^2c^4+G^2 M^2}}{r^2}\\
&=-c_2\frac{\sqrt{3}}{2\sqrt{2}}\dfrac{1}{c^2 }\frac{\sqrt{\dfrac{{\cal J}^2}{M^2}c^2+G^2 M^2}}{r^2}\\
&=-\frac{c_2}{c^2}\frac{\sqrt{3}}{2\sqrt{2}}\frac{\sqrt{\dfrac{{\cal J}^2}{M^2}{c^2}+G^2 M^2}}{r^2}
\end{align}

\begin{figure}[H]
	\centering
	\includegraphics[width=0.45\textwidth]{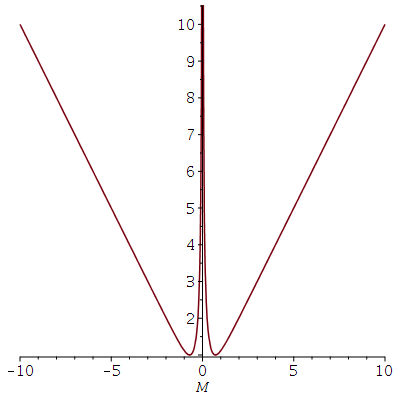}
	\caption{A plot of $\sqrt{\dfrac{{\cal J}^2}{M^2}c^2+G^2 M^2}$ with ${\cal J}=\frac{1}{2}$, $G=1$ and $c=1$}
	\label{twomass}
\end{figure}

It's obvious from the above expression that the mass parameter $m$ can also determine the strength of the electric field, regardless of whether ${\cal J}=0$ or not. The plot of $\sqrt{\dfrac{{\cal J}^2}{M^2}c^2+G^2 M^2}$ in Figure \ref{twomass} shows how two different masses can appear to have the same electric field strength far away from the origin.

It's also obvious from Eq.~(\ref{elecgen}) that $c_2$ being negative or positive makes an important difference. Evidently, from the same equation, $c_2<0$ makes the solution appear to have ``positive'' charge. 

However, Eq.~(\ref{g_tt}) is not so simple. It is clear that $c_1=1$ and $c_1=-1$ both work but the particular choice of sign can impose constraints on the sign of $c_2$ if we are to model the familiar (attractive) Newtonian gravity and vice-versa. Therefore, they must have opposite signs. Else, consider the case when we permanently fix $c_1=+1$ and never allow $c_1=-1$. Then Eq.~(\ref{g_tt}) demands that $c_2=-2$ for attractive gravity. However, by Eq.~(\ref{elecgen}), this would imply that the solution will appear positively charged to a far-off observer, thereby either limiting the scope of the solution or making the negatively ``charged'' solution have a repulsive gravitational field far away from the origin in case of the negatively charged solution.

So, we arrive at these questions:
Can the mass $M$ of an object determine the strength of an electric field? If $M$ is not mass then what is this property ($M$) of the object that determines both its electric and gravitational field strengths?

\section{Discussion}
Although this article does not resolve the questions raised above, we find it pertinent to mention some relevant work from the past. For instance, Kilmister and Stephenson have raised objection to Rainich theory in \cite{Kil54} purely on the grounds that the ``interdependence of gravitational and magnetic fields'' seems to ``conflict with experience (in macroscopic theories)''. In \cite{Klo70}, Klotz and Lynch show an approximate solution to Rainich theory in which they show that the electric field depends on a mass parameter. I am not persuaded that showing a yet unseen or cryptic relationship between mass and charge is proof that the underlying theory is wrong. In \cite{Klo70MNRAS}, the same authors again argue that a spherically symmetric charged body solution in Rainich's Already Unified Theory they constructed will have a smaller size than a neutral one and that this is a ``contradiction'' since they believe that ``electrical repulsion {\it should}\footnote{Emphasis mine. Scientists should not dictate what nature should do.} tend to counteract gravitational attraction to make a charged sphere bigger than an uncharged one''. However, this argument too is unconvincing because it is solely rested upon personal intuition based on everyday experience with charge and gravity and nothing more than that. To show that some object is impossible to exist in nature, one must show a resulting violation of some fundamental law and not everyday physical intuition.

The Schwarzschild-Kerr instanton embbeded solution shown above is in 5 dimensions and no currently accepted physical theory has dimensions other than 4. Therefore, one may reject the 5-dimensional solutions on those grounds alone but that would be insincere, considering theorists are still researching potentially viable string theories in higher dimensions. Another point to note here is that the parameter $M$ does not necessarily have to be interpreted as mass. It can be called by any other name but it is still going to dictate the strength of both the electric and magnetic fields. This strange liaison between mass and electric field should not be brushed under the carpet without rigorously  dismantling either Einstein-Maxwell theory altogether (or at the very least, higher dimensional Einstein-Maxwell theory) or experimentally disproving the theory.

\section{Conclusions}
The article has shown that the notion of electric charge in General Relativity is closely related to mass. Whether this relationship is physically plausible or not remains to be explored. Nonetheless, the present work implores us to find a more fundamental theory of electric charge, just as the Higgs mechanism does for mass.

\bigskip

{\Large Acknowledgements}

This work was partially supported by the Natural Sciences and Engineering Research Council of Canada.


\begin{thebibliography}{99}
	\bibitem{Mis57}
	%
	C.W. Misner, J.A. Wheeeler, ``Classical physics as geometry.'' 
	{\it Ann. Phys.} {\bf 2}, 525 (1957)
	
	\bibitem{Rai25}
	%
	G.Y. Rainich, ``Electrodynamics in the general relativity theory.''
	{\it Trans. Amer. Math. Soc.} {\bf 27}, 106 (1925)
	
	\bibitem{GK2017_1}
	%
	A.M. Ghezelbash, V. Kumar, ``Exact helicoidal and catenoidal solutions in five- and higher-dimensional Einstein-Maxwell theory.'' {\it Phys. Rev. D} {\bf 95}, 124045 (2017)
	
	\bibitem{GK2017_2}
	%
	A.M. Ghezelbash, V. Kumar, ``Exact solutions to Einstein–Maxwell theory on Eguchi–Hanson space.'' {\it Int. J. Mod. Phys. A} {\bf 32}, 1750098 (2017)
	
	\bibitem{Gib77}
	%
	G.W. Gibbons, S.W. Hawking, ``Action integrals and partition functions in quantum gravity.''
	{\it Phys. Rev. D} {\bf 15}, 2752 (1977)
	
	\bibitem{Gro82}
	%
	D.J. Gross, M.J. Perry, L.G. Yaffe, ``Instability of flat space at finite temperature.''
	{\it Phys. Rev. D} {25}, 330  (1982)
	
	\bibitem{Kil54}
	%
	C.W. Kilmister, G. Stephenson, ``An axiomatic criticism of unified field theories\textemdash II.'' 
	{\it Nuovo Cim.} {\bf 11} (Suppl. 1), 118 (1954)
	
	\bibitem{Klo70}
	%
	A.H. Klotz, J. Lynch, ``Is Rainich's theory tenable?.''
	{\it Lett. Nuovo Cim.} {\bf 4} (6), 248 (1970)
	
	\bibitem{Klo70MNRAS}
	%
	A.H. Klotz, J. Lynch, ``A spherically symmetric, charged body in Rainich's Already Unified Field Theory.''
	{\it Mon. Not. R. Astron. Soc.} {\bf 150}, 149 (1970)
	

\end{thebibliography}
\end{document}